\documentclass[twocolumn]{aastex63}
\usepackage{graphicx}	
\usepackage{amsmath}	
\usepackage{amssymb}	
\usepackage{color}
\usepackage{array}
\usepackage{lineno}
\usepackage{bookmark}
\usepackage{subfigure}
\usepackage{booktabs}
\usepackage{threeparttable}

\begin{document}

\title{GRB afterglows with energy injections in AGN accretion disks}

\correspondingauthor{Tong Liu}
\email{tongliu@xmu.edu.cn}

\author[0000-0002-4448-0849]{Bao-Quan Huang}
\affiliation{Department of Astronomy, Xiamen University, Xiamen, Fujian 361005, China}

\author[0000-0001-8678-6291]{Tong Liu}
\affiliation{Department of Astronomy, Xiamen University, Xiamen, Fujian 361005, China}

\author{Xiao-Yan Li}
\affiliation{Department of Astronomy, Xiamen University, Xiamen, Fujian 361005, China}

\author[0000-0002-9130-2586]{Yun-Feng Wei}
\affiliation{Department of Astronomy, Xiamen University, Xiamen, Fujian 361005, China}

\begin{abstract}
Active galactic nucleus (AGN) disks are widely considered potential hosts for various high-energy transients, including gamma-ray bursts (GRBs). The reactivation of GRB central engines can provide additional energy to shocks formed during the interaction of the initially ejected GRB jets with the circumburst material, commonly referred to as energy injections. In this paper, we study GRBs occurring in AGN disks within the context of energy injections. We adopt the standard external forward shock (EFS) model and consider both short- and long-duration GRB scenarios. Light curves for two types of radiation, namely the radiation from the heated disk material (RHDM) and GRB afterglows, are computed. We find that the energy injection facilitates the EFS to break out from the photosphere of the low-density AGN disk at relativistic velocity. Moreover, the energy injection almost does not affect the RHDM but significantly enhances the peak flux of the GRB afterglows.
\end{abstract}


\section{Introduction}

The accretion disks in active galactic nuclei (AGNs) have been extensively investigated since the discovery of AGNs in the last century \citep[see reviews by][]{Pringle1981,Kato2008book,Abramowicz2013}. It is commonly accepted that AGN disks harbor numerous stars and compact objects. These stars and compact objects can either form within the disks \citep[e.g.,][]{Goodman2003,Artale2019,Dittmann2020,Cantiello2021,Fan2023} or be captured from outside \citep[e.g.,][]{Artymowicz1993,Fabj2020,MacLeod2020,Nasim2023,Generozov2023,Wang2024}, and some of them probably exist in binary systems \citep[e.g.,][]{Baruteau2011,Pfuhl2014,Bartos2017}.
The collapses, explosions, or tidal disruptions of the stars, as well as the collisions/mergers between or among the compact objects, are generally followed by various transient events across electromagnetic wavelengths, such as gamma-ray bursts \citep[GRB, e.g.,][]{Perna2021,Zhu2021,Yuan2022,Lazzati2022,Qi2022,Wang2022,Tagawa2023,Ray2023,Lazzati2023,Kathirgamaraju2023}.

GRBs stand as the most brilliantly luminous explosions in the Universe. They are broadly categorized into two groups: short- and long-duration GRBs (SGRBs and LGRBs), where SGRBs are generally considered to be associated with the mergers of compact objects, whereas LGRBs are linked to the collapses of massive stars \citep[see reviews by][]{Woosley2006,Nakar2007,Zhang2018book}. Additionally, they are widely believed to be generated by ultrarelativistic jets originating from their central engines, which could be either a stellar-mass black hole (BH) surrounded by a hyperaccretion disk \citep[for reviews, see][]{Liu2017} or a massive millisecond magnetar \citep[e.g.,][]{Duncan1992,Usov1992,Dai1998b,Zhang2001}.

In general, after the prompt $\gamma$-ray emission, enduring multiband afterglows arise from synchrotron radiation, which stems from electrons accelerated by external forward shocks (EFSs) formed as the jets interplay with the circumburst material \citep[e.g.,][]{Zhang2006}. Observationally, a substantial portion of X-ray afterglows exhibit single or double plateaus \citep[e.g.,][]{Li2018,Hou2021,Yi2022} and single or multiple flares \citep[e.g.,][]{Mu2016,Liu2019} in light curves. Regarding these phenomena, primary explanations involve the reactivation of GRB central engines. This reactivation can directly contribute to the flares \citep[e.g.,][]{Burrows2005,Dai2006}, or provides energy injections into EFSs to indirectly form the plateau phase \citep[e.g.,][]{Zhang2001,Li2016,Hou2018,Lin2018,Huang2021}. According to the statistical analyses, we consider that most of X-ray plateaus and flares might have the same physical origin but appear with different features \citep[e.g.,][]{Yi2022}.

Recently, significant attention has been directed towards GRBs occurring in AGN disks. Unlike typical environments, GRBs within AGN disks can experience diffuse emergence due to the high densities of disk material \citep{Perna2021}. Additionally, the diffuse afterglows are fairly weak and peak at very late stages \citep{Wang2022}. To comprehend the role of reactivated GRB central engines within AGN disks, we investigate the effects of energy injections resulting from their reactivation on the dynamical evolution of EFSs and the characteristics of relevant radiations in this work.

The paper is structured as follows. In Section 2, we present the AGN disk model, the EFS-dynamics model, and relevant radiation mechanism. The main results are outlined in Section 3, and the conclusions and discussion are made in Section 4.

\begin{figure}
\centering
\includegraphics[width=1.0\linewidth]{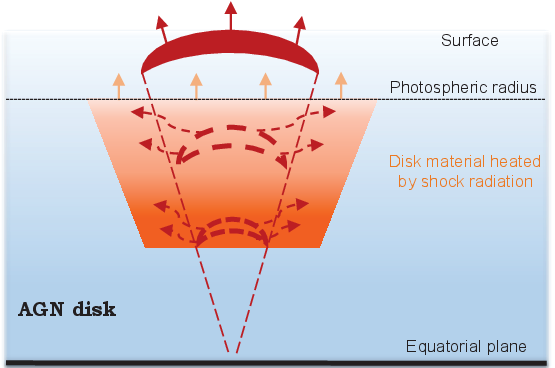}
\caption{Illustration of a GRB jet propagating in an AGN disk.}
\label{Figure1}
\end{figure}

\section{Method}

We investigate the physical processes of EFSs with energy injections propagating in AGN disks. We assume EFSs are generated during the initially ejected GRB jets interact with the circumburst medium. Here, GRB jets are assumed to be launched close to the equatorial plane of the disks and to move vertically upward, as depicted in Figure \ref{Figure1}. During the propagation of EFSs in the disks, on the one hand, the reactivation of GRB central engines pushes out new jets; these new jets catch up with EFSs, continuously providing them with energy injections. On the other hand, the radiation emitted by EFSs is significantly scattered and absorbed by the high-density material of the disks. For simplicity, we assume that the material eventually absorbs all radiation from EFSs in regions where the optical depth exceeds one, leading to heating and subsequent thermal emission, hereafter referred to as radiation from the heated disk material (RHDM). Conversely, we neglect the material's effects on photon propagation in regions where the optical depth is less than one. Thereby, once EFSs penetrate through the photosphere of the disks, GRB afterglows become transparent. Note that here the EFSs involve the initially ejected GRB jets.

\subsection{AGN accretion disk model}

We adopt the accretion disk model proposed by \cite{Sirko2003}. This disk model considers the effect of self-gravity and provides an accurate description for the radial structure from inner regions out to $\sim 2\times10^5{R_{\rm g}}$, where $R_{\rm g}=2GM_{\rm BH}/c^2$ is the Schwarzschild radius with $G$ being the gravitational constant, $M_{\rm BH}$ being the mass of BHs, and $c$ being the speed of light. Additionally, within this disk model, the supermassive BH (SMBH) spin is not involved, and the outer part of the accretion disks belongs to the non-ionizing region due to significantly low temperatures. The SMBH masses are set to $M_{\rm BH}=10^6$, $10^7$, and $10^8 \, M_\odot$, the SMBH accretion efficiency $l_{\rm Edd}=0.5$ ($l_{\rm Edd}=0.1 \dot{M} c^2/L_{\rm Edd}$, where $\dot{M}$ and $L_{\rm Edd}$ are the accretion rate and the Eddington luminosity, respectively), and the viscosity parameter $\alpha=0.01$.

For the vertical structure, we utilize a Gaussian function to depict the density distribution, which is written as follows
\begin{equation}\label{eqn:1}
\rho_{\rm disk}(R_r,R_z)=\rho_{\rm 0}(R_r)\exp \left[{-\frac{R_z^2}{2H(R_r)^2}} \right],
\end{equation}
where $R_r$ represents the radial radius, $R_z$ stands for the vertical distance from the equatorial plane, $\rho_{\rm 0}(R_r)$ denotes the density at the equatorial plane, and $H(R_r)$ signifies the half-thickness of accretion disks. Further, the photospheric radius of accretion disks, $R_{\rm ph,disk}$, defined as the location where the optical depth is one from the surface of accretion disks inwards, can be calculated by
\begin{align}\label{eqn:2}
\tau(R_r)=\int_{R_{\rm ph,disk}}^{H}\kappa_{0}(R_r)\rho_{\rm disk}(R_r,R_z)\,dR_z=1,
\end{align}
where $\kappa_{0}(R_r)$ is the opacity of accretion disks.

\subsection{Dynamical evolution of EFSs with energy injections}

\begin{figure}
\includegraphics[width=0.95\linewidth]{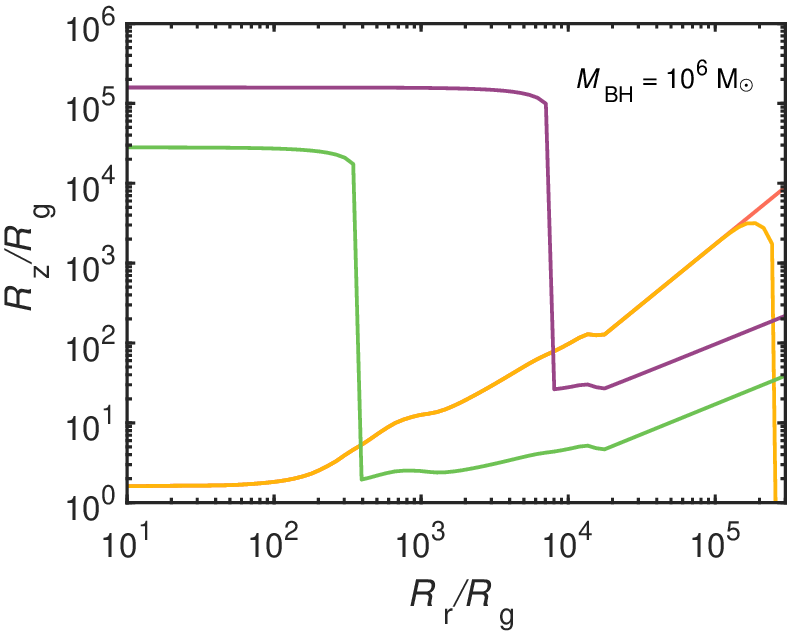}
\includegraphics[width=0.95\linewidth]{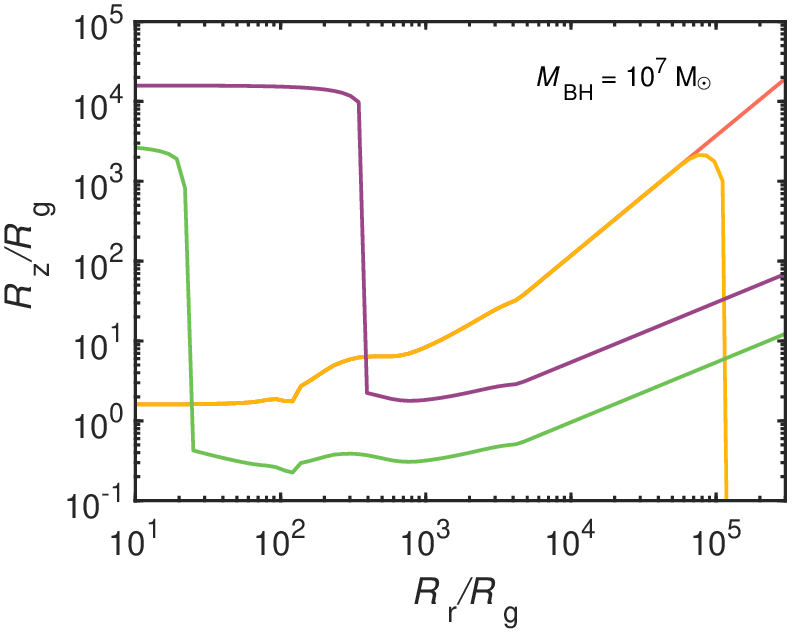}
\includegraphics[width=0.95\linewidth]{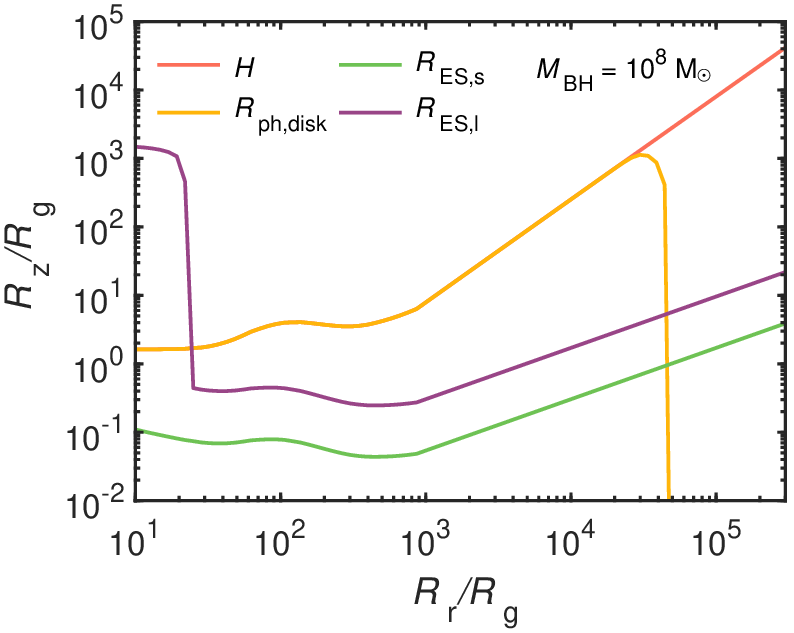}
\caption{The initial radii of EFSs in SGRBs ($R_{\rm ES,s}$) and LGRBs ($R_{\rm ES,l}$), the photospheric radius of AGN disks ($R_{\rm ph,disk}$), and the disk disk half-thickness ($H$) evolving with radial locations for different SMBH masses, including $10^{6}\,M_\odot$ (top panel), $10^{7}\,M_\odot$ (middle panel), and $10^{8}\,M_\odot$ (bottom panel).}
\label{Figure2}
\end{figure}

Assuming the initially ejected GRB jets to be thick shells, the initial position of EFSs can be written as \citep{Sari1995}
\begin{equation}\label{eqn:3}
R_{\rm ES}=R_{\rm Sedov}^{3/4}\Delta^{1/4},
\end{equation}
where $R_{\rm Sedov}$ represents the Sedov radius defined by the integral
\begin{equation}\label{eqn:4}
\int_{R_0}^{R_{\rm Sedov}}4\pi R^2 \rho c^2\,dR=E_{\rm iso},
\end{equation}
with ${R_0}$ and $E_{\rm iso}$ being the initial position and isotropic energy of the initially ejected GRB jets, and $\Delta=cT$ is the lab-frame jet width with $T$ being the duration of GRBs. $\rho$ is the density of the circumburst medium with $\rho=\rho_{\rm disk}$ inside the disks and $\rho=n_{\rm ISM} m_{\rm p}$ outside the disks, where $n_{\rm ISM}$ is the number density of the interstellar medium (ISM) and $m_{\rm p}$ is the proton mass. In this work, we consider both SGRB and LGRB situations, and denote the corresponding initial radii of EFSs as $R_{\rm ES,s}$ and $R_{\rm ES,l}$. $R$ represents the distance from the central engines. Here the jet direction along the $z$ axis is considered. For SGRBs, we set ${R_0}=10^7 \,{\rm cm}$, $E_{\rm iso}=10^{51} \,{\rm erg}$, and $T=1 \,{\rm s}$. Meanwhile, for LGRBs, we set ${R_0}=10^{10} \,{\rm cm}$, $E_{\rm iso}=10^{53} \,{\rm erg}$, and $T=10 \,{\rm s}$.

We use an approximate dynamical evolution model to discuss the evolution of EFSs, which contains differential equations listed as follows \citep{Huang1999},
\begin{equation}\label{eqn:5}
\frac{d \Gamma}{dt}=\frac{-(\Gamma^{2}-1)\frac{dm}{dt}+\frac{1}{c^2}\frac{dE_{\rm in}}{dt}}{M_{\rm j}+\epsilon m + 2(1- \epsilon ) \Gamma m},
\end{equation}
\begin{equation}\label{eqn:6}
\frac{dm}{dt}=2 \pi (1-\cos\theta_{\rm jet}) R^{2} \rho \frac{dR}{dt},
\end{equation}
\begin{equation}\label{eqn:7}
\frac{dR}{dt} =\frac{c\beta}{(1-\beta\cos\theta_{\rm v})(1+z)}.
\end{equation}
Here, $\Gamma$, $m$, $\epsilon$, and $\theta_{\rm jet}$ are the Lorentz factor, swept mass of the external medium, radiation efficiency of EFSs, and jet-half opening angle, respectively. $M_{\rm j}=E_{\rm k,iso}(1-\cos\theta_{\rm jet})/2(\Gamma_0-1)c^2$ is the mass of the initially ejected GRB jets, where $E_{\rm k,iso}$ is the isotropic kinetic energy. $\beta =\sqrt{1-1/\Gamma^2}$ is the EFS velocity. $t$ is the time measured in the observer's frame. $\theta_{\rm v}$ is the viewing angle, which is the angle between the jet axis and the line of sight, and it is set to $0^{\circ}$ here, i.e., on-axis viewing. $z$ is the redshift and is set to 0.5. In addition, the functional form of $dE_{\rm in}/dt$ can be simply described as \citep{Lin2018}
\begin{equation}\label{eqn:8}
  {\frac{dE_{\rm in}}{dt}}=
  \left\{
  \begin{array}{ll}
  \frac{P_{\rm iso,in}(1-\cos\theta_{\rm jet})}{2},\quad T_{\rm st,in} < t < T_{\rm st,in}+T_{\rm st,du},\\
  0, \quad \quad \quad \quad \quad \quad \quad \quad \quad \quad {\rm others},
  \end{array}
  \right.
\end{equation}
where $P_{\rm iso,in}$, $T_{\rm st,in}$, and $T_{\rm st,du}$ are the isotropic power, start time, and duration of energy injections, respectively. For simplicity, in this work we only consider a constant injected power, namely $P_{\rm iso,in}={\rm constant}$.

\begin{table*}[!t]
    \footnotesize
    \centering
    \caption{The parameter values for selected 9 cases. These parameters are the SMBH masses, the radial locations from the centers, the initial radii of EFSs in SGRBs, the initial radii of EFSs in LGRBs, the photospheric radii of AGN disks, and the half-thickness of AGN disks, respectively.}
    \renewcommand\arraystretch{1.2}
    \setlength{\tabcolsep}{1.0mm}
    \begin{threeparttable}
    \begin{tabular}{ccccccc}
    \toprule[0.8pt]
    $\rm Cases$   &\quad \quad  $M_{\rm BH}\,(M_\odot)$   &\quad \quad  $R_r\,(R_{\rm g})$   &\quad \quad  $R_{\rm ES,s}\,(R_{\rm g})$   &\quad \quad     $R_{\rm ES,l}\,(R_{\rm g})$   &  \quad \quad     $R_{\rm ph,disk}\,(R_{\rm g})$   &  \quad \quad     $H\,(R_{\rm g})$ \\
    \midrule[0.6pt]
    $\text{1}$   &\quad \quad $10^6$   &\quad \quad $10^3$   &\quad \quad $2.49\times 10^{0}$    &\quad \quad $1.563\times 10^{5}$   &\quad \quad $1.262\times 10^{1}$  &\quad \quad $1.262\times 10^{1}$ \\

    $\text{2}$   &\quad \quad $10^6$   &\quad \quad $10^4$   &\quad \quad $4.701\times 10^{0}$   &\quad \quad $2.787\times 10^{1}$   &\quad \quad $9.703\times 10^{1}$  &\quad \quad $9.705\times 10^{1}$ \\

    $\text{3}$   &\quad \quad  $10^6$  &\quad \quad $10^5$   &\quad \quad $1.708\times 10^{1}$   &\quad \quad $9.652\times 10^{1}$   &\quad \quad $1.721\times 10^{3}$   &\quad \quad $1.726\times 10^{3}$\\

    $\text{4}$   &\quad \quad  $10^7$  &\quad \quad $10^3$  &\quad \quad $3.181\times 10^{-1}$    &\quad \quad $1.831\times 10^{0}$  &\quad \quad $8.377\times 10^{0}$  &\quad \quad $8.379\times 10^{0}$ \\

    $\text{5}$   &\quad \quad $10^7$   &\quad \quad $10^4$  &\quad \quad $9.606\times 10^{-1}$   &\quad \quad $5.414\times 10^{0}$   &\quad \quad $1.171\times 10^{2}$  &\quad \quad $1.171 \times 10^{2}$\\

    $\text{6}$   &\quad \quad $10^7$   &\quad \quad $10^5$  &\quad \quad $5.401\times 10^{0}$   &\quad \quad $3.038\times 10^{1}$   &\quad \quad     $1.648\times 10^{3}$  &\quad \quad $3.718 \times 10^{3}$\\

    $\text{7}$   &\quad \quad $10^8$  &\quad \quad $10^3$  &\quad \quad $5.405\times 10^{-2}$   &\quad \quad $3.043\times 10^{-1}$    &\quad \quad       $7.893\times10^{0}$   &\quad \quad $7.895 \times 10^{0}$\\

    $\text{8}$   &\quad \quad $10^8$  &\quad \quad $10^4$  &\quad \quad $3.037 \times 10^{-1}$  &\quad \quad $1.708 \times 10^{0}$   &\quad \quad      $2.522\times 10^{2}$  &\quad \quad $2.524 \times 10^{2}$\\

    $\text{9}$   &\quad \quad $10^8$  &\quad \quad $10^5$  &\quad \quad $1.708\times 10^{0}$   &\quad \quad $9.605\times 10^{0}$    &\quad \quad      $0.0$  &\quad \quad $8.010\times 10^{3}$ \\
    \bottomrule[0.6pt]
    \label{table1}
\end{tabular}
\end{threeparttable}
\end{table*}

\subsection{RHDM}

We assume that the disk material begins to be heated as EFSs form. Therefore, the lower position of the region of the heated disk material is at the initial radius of EFSs, and the upper position of this region is set to the photospheric radius of the disks, as shown in Figure \ref{Figure1}. To calculate the radiation from this region, we divide it into a series of shell layers from bottom to top, which is similar to the treatment of the kilonova or supernova radiation calculations \citep[e.g.,][]{Chen2022,Qi2022b,Li2023}. For each shell layer, the evolution of the internal energy can be written as
\begin{equation}\label{eqn:9}
  {\frac{dE_{i}}{dt_{\rm b}}}=
  \left\{
  \begin{array}{lll}
  0,                 \quad \quad \quad \quad \quad R(t_{\rm b}) < R_{\rm i,1},\\
  L_{\rm inj}-L_{i}, \quad R_{\rm i,1} < R(t_{\rm b}) < R_{\rm i,2},\\
  -L_{i},            \quad \quad \quad \quad R_{\rm i,2} < R(t_{\rm b}).\\
  \end{array}
  \right.
\end{equation}
Here the subscript $i$ represents the $i$th layer, so $E_{i}$ is the internal energy of the $i$th layer, $L_{i}$ is the observed luminosity contributed by the $i$th layer, $R_{\rm i,1}$ is the lower radius of the $i$th layer, and $R_{\rm i,2}$ is the upper radius of the $i$th layer. $t_{\rm b}$ is the time measured in the lab frame, and the corresponding observer time is equal to $t_{\rm b}(1+z)$ for the shell layers. Note that the observer time, before EFSs form, is approximately equal to $R_{\rm ES}(1+z)/2\Gamma_0^2c$. $R(t_{\rm b})$ is the radius of EFSs, and $L_{\rm inj}$ is the luminosity of the radiation escaping from EFSs, which can be expressed as \citep[e.g.,][]{Panaitescu1998,Piran1999}
\begin{equation}\label{eqn:10}
L_{\rm inj}=\epsilon (\Gamma-1)c^2\frac{dm}{dt_{\rm b}}.
\end{equation}

The observed luminosity contributed by the $i$th layer can be obtained by the following formula,
\begin{align}\label{eqn:11}
L_i=\frac{E_{i}}{{\rm max}[t_{{\rm d},i},t_{{\rm lc},i}]},
\end{align}
where $t_{{\rm lc},i}=R_{i}/c$ represents the light-crossing time, and  $t_{{\rm d},i}$ is the radiation diffusion timescale, which can be written as
\begin{align}\label{eqn:12}
t_{{\rm d},i}=\frac{3\kappa_0}{\Omega_{\rm c} c}\sum_{j=i}^{n}\frac{m_{j}}{R_{j}},
\end{align}
where $m_{j}$ and $R_{j}$ are the mass and the middle radius of the $j$th layer. Here, $m_{j}=\Omega_{\rm c} (R_{j,2}^3-R_{j,1}^3) \rho_{\rm disk}/3$, and $\Omega_{\rm c}=2 \pi (1-\cos\theta_{\rm c})$, with $\theta_{\rm c}$ being the half opening angle of the heated disk metarial, which is set as $10^{\circ}$.

The overall bolometric luminosity of the heated disk metarial finally can be calculated by accumulating the contributions from each shell layer,
\begin{equation}\label{eqn:13}
L_{\rm ph}=\sum_{i=1}^{n} L_i.
\end{equation}
Assuming that this radiation approximately satisfies the blackbody spectrum and is emitted from the photosphere of the disks, the effective temperature can be expressed as
\begin{equation}\label{eqn:14}
T_{\rm eff}=\left(\frac{L_{\rm ph}}{\Omega_{\rm c} R_{\rm ph,disk}^2 \sigma_{\rm SB}} \right)^{1/4},
\end{equation}
where $\sigma_{\rm SB}$ is the Stephan-Boltzmann constant. In addition, the flux density can be written as
\begin{equation}\label{eqn:15}
F_{\nu_{\rm obs}} (t)=\frac{2 \pi h \nu^3}{c^2}\frac{(1+z)}{\exp(h\nu/kT_{\rm eff})-1}\frac{\Omega_{\rm c}R_{\rm ph,disk}^2}{4 \pi D_{\rm L}^2},
\end{equation}
where $\nu$ represents the local frequency, related to the observer frequency $\nu_{\rm obs}$ by a factor of $(1+z)$, $k$ denotes the Boltzmann constant, and $D_{\rm L}$ is the luminosity distance in the standard $\Lambda$CDM cosmology model ($\Omega_M=0.27$, $\Omega_\Lambda=0.73$, and $H_0=71~\rm km~s^{-1}~Mpc^{-1}$).

\subsection{Synchrotron radiation}

After EFSs break out of disk photospheres, afterglows become transparent. At this stage, we calculate the synchrotron radiation emitted by the electrons of the circumburst medium accelerated by the EFSs.

Based on the above-mentioned EFS dynamics, the evolution of the shock-accelerated electrons is obtained by the following formula \citep{Fan2008},
\begin{equation}\label{eqn:16}
\frac{\partial }{{\partial R}}\left(\frac{{dN'_{\rm{e}}}}{{d\gamma _{\rm{e}}^\prime }}\right) + \frac{\partial }{{\partial \gamma _{\rm{e}}^\prime }}\left( \frac{d \gamma _{\rm{e}}^\prime}{{dR}}\frac{{dN'_{\rm{e}}}}{{d\gamma _{\rm{e}}^\prime }}\right) = {Q},
\end{equation}
where $dN'_{\rm{e}}/d\gamma _{\rm{e}}^\prime $ is the instantaneous electron energy spectrum, $\gamma^\prime_{\rm e}$ is the Lorentz factor of the shock-accelerated electrons, $d\gamma _{\rm{e}}^\prime/dR$ is the cooling term of electrons with the Lorentz factor $\gamma^\prime_{\rm e}$, and $Q$ is the injection term for newly shocked-accelerated electrons. Since the shock-accelerated electrons are assumed to follow a power-law distribution, one can express $Q = \bar{K} \gamma _{\rm{e}}'^{-p}$, with $\bar{K}\approx 2 \pi (1-\cos\theta_{\rm jet})(p-1)R^2n_{\rm}\gamma_{\rm e,min}'^{p-1}$, for $\gamma^\prime_{\rm e,min}\leq\gamma^\prime_{\rm e}\leq\gamma^\prime_{\rm e,max}$, where $p$ ($> 2$), $\gamma^\prime_{\rm e,min}$, and $\gamma^\prime_{\rm e,max}$ represent the power-law index, the minimum Lorentz factor, and the maximum Lorentz factor of the shock-accelerated electrons, respectively. Here, the minimum and maximum Lorentz factor can be expressed as $\gamma_{\rm e,min}'=\epsilon_{\rm e}(\Gamma-1)(p-2) m_{\rm p}/(p-1)m_{\rm e}+1$ and $\gamma^\prime_{\rm e,max}=\sqrt{{9m_{\rm e}^{2}c^{4}}/{8B'e^3(1+Y)}}$, respectively, with $B^\prime=\sqrt{32 \pi \Gamma (\Gamma-1) \rho \epsilon_{B}c^2}$ \citep{Kumar2012}, where $\epsilon_{\rm e}$ is the fraction of the thermal energy density that is shared by accelerated elections, $\epsilon_{B}$ is the fraction of the thermal energy density that is shared by magnetic filed, $e$ is the electron charge, $m_{\rm e}$ is the electron masses, and $Y$ is the Compton parameter \citep{Fan2006}. Additionally, the radiation efficiency of EFSs can be expressed as $\epsilon=\epsilon_{\rm rad}\epsilon_{\rm e}$, with $\epsilon_{\rm rad}={\rm min}[1,(\gamma_{\rm e,min}'/\gamma_{\rm c}')^{(p-2)}]$, where $\gamma_{\rm c}$ is the efficient cooling Lorentz factor of electrons \citep{Sari2001}. Note that quantities with a superscript accent sign are defined in the comoving frame of EFSs.

The spectral power of synchrotron radiation at frequency $\nu'$ can be expressed as \citep{Rybicki1979}
\begin{equation} \label{eqn:17}
P'_{\rm syn}({\nu'})=\int\nolimits_{\gamma'_{\rm e,min}}^{\gamma'_{\rm e,max}} P'_{\rm e}({\nu'}, \gamma'_{\rm e}) \frac{dN'_{\rm e}}{d\gamma'_{\rm e}} d\gamma'_{\rm e},
\end{equation}
where
\begin{equation}\label{eqn:18}
P'_{\rm e}({\nu'}, \gamma'_{\rm e})=\frac{\sqrt{3} e^3 B'}{m_{\rm e}c^2} F \bigg(\frac{\nu'}{\nu'_{\rm c}} \bigg),
\end{equation}
represents the power of a single election with the Lorentz factor $\gamma'_{\rm e}$. Here, $\nu_{\rm c}'=3 e B' \gamma_{\rm e}'^{2}/4 \pi m_{\rm e}c$, $F({\nu'}/{\nu_{\rm c}'} )=(\nu'/\nu_{\rm c}') \int\nolimits_{\nu'/\nu_{\rm c}'}^{+ \infty} K_{5/3}(x) dx$, and $K_{5/3}(x)$ is a modified Bessel function of order 5/3. Since elections with $\gamma'_{\rm e}<3$ are thought to be difficult to produce synchrotron radiation, the lower integration limit in Equation (\ref{eqn:17}) is set to 3 if $\gamma'_{\rm e,min} < 3$. In addition, the self-absorption effect of synchrotron radiation is taken into account in this work, and the corresponding optical depth is given by \citep{Rybicki1979}
\begin{equation}\label{eqn:19}
\tau_{\rm tot} ({\nu'})=\frac{1}{8 \pi \nu'^2 m_{\rm e} f} \int P'_{\rm e}({\nu'}, \gamma'_{\rm e}) \gamma'^2_{\rm e} \frac{d}{d\gamma'_{\rm e}} \left(\frac{dN'_{\rm e}/d\gamma'_{\rm e}}{\gamma'^2_{\rm e}} \right) d\gamma'_{\rm e}
\end{equation}
with $f=2 \pi (1-\cos\theta_{\rm jet}) R^2$. Consequently, the spectral power of synchrotron radiation considering the self-absorption effect can be written as
\begin{align}\label{eqn:20}
P'_{\rm syn,sa}({\nu'}) = \int_{0}^{\tau_{\rm tot}} \frac{P'_{\rm syn}({\nu'})}{\tau_{\rm tot}({\nu'})} \exp{(-\tau)} d\tau.
\end{align}

We consider a top-hat jet and divide it into a number of emitters. Therefore, the observed flux density is calculated by accumulating the flux density of the emitters at the same observer time, which can be derived as \citep[e.g.,][]{Granot1999}
\begin{align}\label{eqn:21}
F_{\nu_{\rm obs}}=\frac{1+z}{4\pi D_{\rm L}^2}
{\int}\kern-15pt\int\limits_{(\rm EATS)} \frac{P'_{\rm syn,sa}(\nu')}{2 \pi (1-\cos\theta_{\rm jet})} {D^3} d\Omega,
\end{align}
where ``EATS'' refers to the equal-arrival time surface corresponding to the same observer time, $\nu^{\prime}=(1+z)\nu_{\rm obs}/D$, and $D$ represents the Doppler factor of the emitters.

\section{Results}

\begin{table}
    \footnotesize
    \centering
    \caption{The chosen values of EFS parameters for both SGRBs and LGRBs.}
    \renewcommand\arraystretch{1.2}    
    \setlength{\tabcolsep}{1.0mm}    
    \begin{threeparttable}
    \begin{tabular}{lccc}
    \toprule[0.8pt]
     Parameter               &\quad \quad Unit                 &\quad \quad SGRB               &\quad \quad LGRB\\
    \midrule[0.6pt]
    $E_{\rm k,iso}$           &\quad \quad ${\rm erg}$          &\quad \quad $10^{51}$          &\quad \quad $10^{53}$\\
    $\Gamma_{\rm 0}$          &                                 &\quad \quad $2\times 10^{2}$   &\quad \quad $2\times 10^{2}$\\
    $\epsilon_{\rm e}$        &                                 &\quad \quad $10^{-1}$          &\quad \quad $10^{-1}$\\
    $\epsilon_{\rm B}$        &                                 &\quad \quad $10^{-3}$          &\quad \quad $10^{-3}$\\
    $n_{\rm ISM}$             &\quad \quad ${\rm cm}^{-3}$      &\quad \quad $1$                &\quad \quad $1$\\
    $p$                       &                                 &\quad \quad $2.5$              &\quad \quad $2.5$\\
    $\theta_{\rm jet}$          &\quad \quad degree               &\quad \quad $5$                &\quad \quad $5$\\
    $T_{\rm st,in}$           &\quad \quad ${\rm s}$            &\quad \quad $10^{1}$           &\quad \quad $10^{1}$\\
    $T_{\rm du,in}$          &\quad \quad ${\rm s}$            &\quad \quad $2 \times 10^{2}$           &\quad \quad $2 \times 10^{3}$\\
    $P_{\rm iso,in}$         &\quad \quad ${\rm erg/s}$        &\quad \quad $10^{49}$          &\quad \quad $10^{51}$\\
    \bottomrule[0.8pt]
    \label{table2}
    \end{tabular}
    \end{threeparttable}
\end{table}

\begin{figure*}[!t]
\centering
\includegraphics[width=1.0\linewidth]{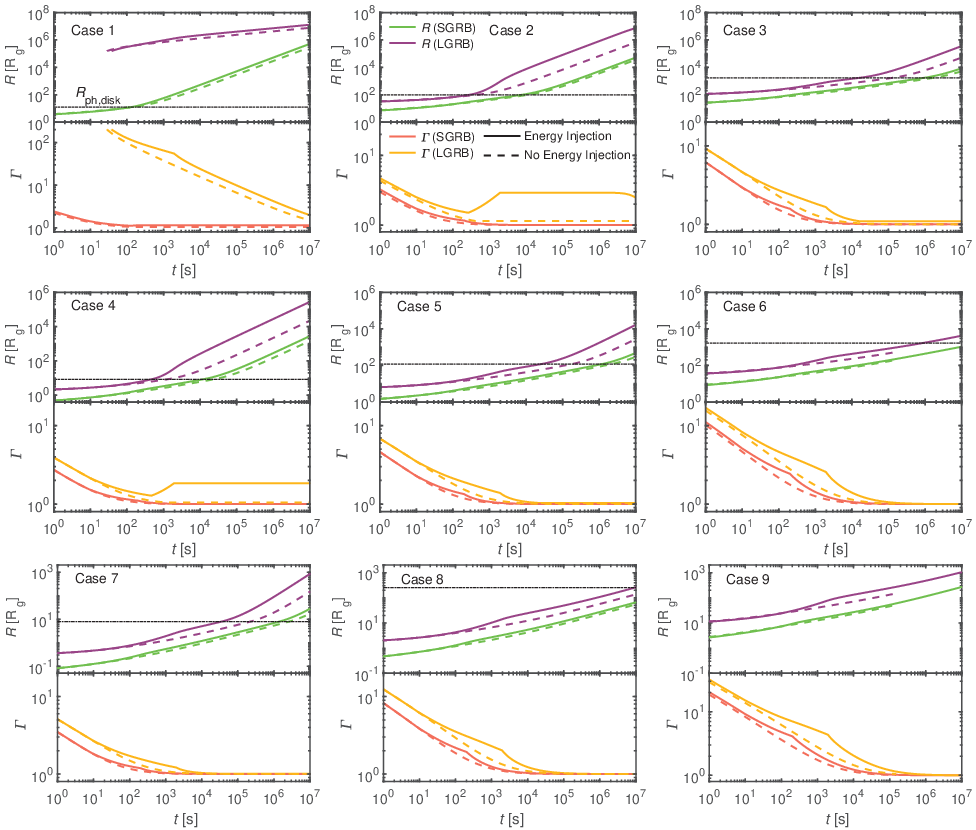}
\caption{$\emph{R}$ and $\Gamma$ of evolving EFSs in 9 cases. In each panel, green (purple) lines represent the evolution of $R$ in SGRBs (LGRBs), and the red (yellow) lines represent the evolution of $\Gamma$ in SGRBs (LGRBs), where solid (dashed) lines denote the results with (without) energy injections. In addition, the black dot-dash lines show the photospheric radius of AGN disks ($R_{\rm ph,disk}$).}
\label{Figure3}
\end{figure*}

Figure \ref{Figure2} illustrates the evolutions of the initial EFS radii, disk photospheric radius, and disk half-thickness concerning the radial radius. The top, middle, and bottom panels present the results corresponding to SMBH masses of $M_{\rm BH}=10^6$, $10^7$, and $10^8 \, M_\odot$, respectively. In Figure \ref{Figure2}, it is evident that EFSs in LGRBs exhibit larger initial radii compared to those in SGRBs. Additionally, at smaller radial radii, the initial EFS radii surpass the disk half-thickness, indicating the formation of EFSs outside the disks. However, with the increase in radial radius, the disk half-thickness undergoes a significant increase. At a specific location, the initial EFS radii suddenly become smaller than the disk half-thickness, implying the inception of EFS formation within the disks from that radial location onwards. Moreover, irrespective of the SMBH masses, the photospheric radius of the disks (yellow lines) essentially aligns with the disk half-thickness (red lines) within the region of $R_{\rm r}<10^5 R_{\rm g}$. This alignment suggests considerable opacity of the disks in this area, with the photospheres almost coinciding with the disk surfaces. Beyond this region, however, the disks quickly become completely transparent. In this study, to encompass all scenarios, namely EFSs forming in both the optically thin and thick regions of the disks as well as outside the disks, while minimizing computational efforts, we adopt three radial radii: $R_r=10^3$, $10^4$, and $10^5 R_{\rm g}$ for each SMBH mass, according to the results of Figure \ref{Figure2}. Consequently, our investigation encompasses 9 cases, as outlined in Table \ref{table1}. Our focus lies on studying the EFS dynamical evolution and the characteristics of relevant radiations across these 9 cases, with particular emphasis on the effects of energy injections on the EFSs.

In general, there are two forms of continuous energy injection into EFSs, one due to a long-lived central engine \citep[e.g.,][]{Da1998a,Zhang2001,Kumar2008,Huang2021} and the other due to a stratification of the ejecta Lorentz factor \citep[e.g.,][]{Rees1998,Sari2000,Granot2003}. Regardless of whether the progenitor is the merger of compact binaries or the collapse of massive stars, either a rapidly rotating stellar-mass BH or a massive millisecond magnetar is involved in the center. In the case of BH engines, continuous energy injections are attributed to material accretion, and in the case of magnetar engines, continuous energy injections are supplied by the Poynting flux/electron-positron wind. The energy of the magnetar wind cannot exceed the maximum rotational energy of $\sim 2 \times10^{52} \, {\rm erg}$, but this limit does not exist in the BH accretion scenario. Compared to the normal circumstellar environment, in AGN disks, more massive progenitor stars prefer to produce BHs and more massive fallback accretion material should prolong the duration of energy injections and improve the total energy injected. Thus, continuous energy injections in the BH accretion scenario are effective in helping EFSs to break out of the disk photospheres. However, to be conservative, we do not consider specific scenarios and only base on the results of the observed sample statistics to set $E_{\rm k,iso}=10^{51} \,{\rm erg}$, $P_{\rm iso,in}=10^{49} \,{\rm erg \,s^{-1}}$, and $T_{\rm du,in}=2\times10^2 \, {\rm s}$ for SGRBs, and $E_{\rm k,iso}=10^{53} \,{\rm erg}$, $P_{\rm iso,in}=10^{51} \,{\rm erg \,s^{-1}}$, and $T_{\rm du,in}=2\times10^3 \, {\rm s}$ for LGRBs, additionally $T_{\rm st,in}=10 \, {\rm s}$ for both. The detailed values of EFS parameters are listed in Table \ref{table2}.

\begin{figure*}[!t]
\centering
\includegraphics[width=1.0\linewidth]{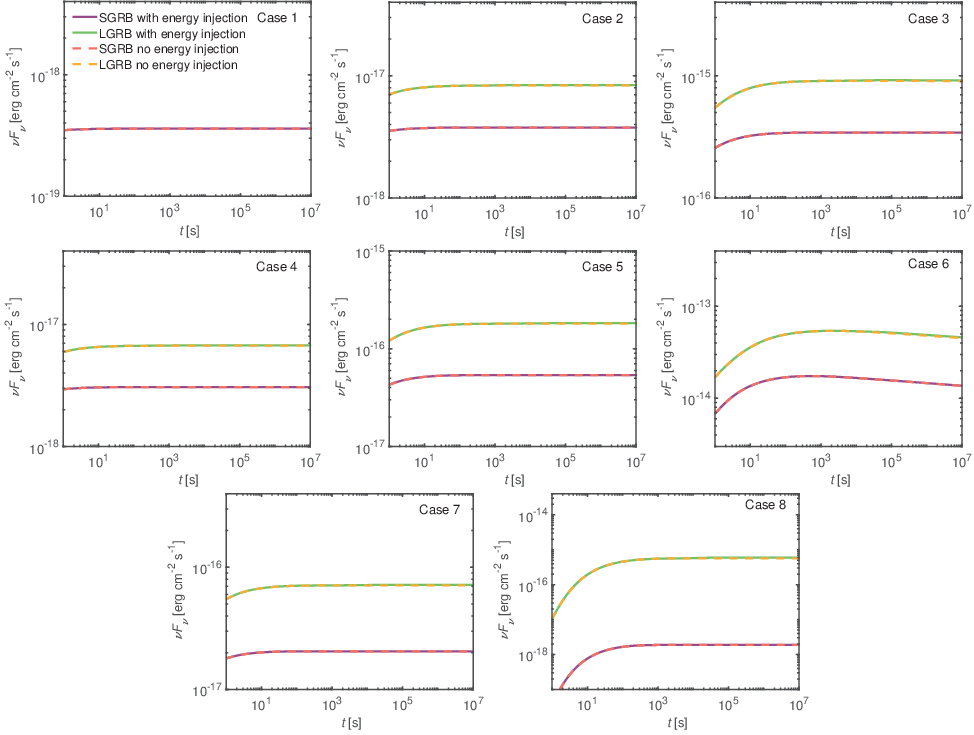}

\caption{Optical light curves of RHDM. The purple (green) solid lines represent the results of SGRBs (LGRBs) with the energy injections, and the red (yellow) dashed lines denote the results of SGRBs (LGRBs) without the energy injections.}
\label{Figure4}
\end{figure*}

\subsection{Evolutions of EFSs}

Figure \ref{Figure3} displays the evolutions of $R$ and $\Gamma$ across observer time in 9 cases, with each panel corresponding to a distinct case. Within each panel, two sections are present: the upper section delineates the $R$ evolutions in SGRBs (green lines) and LGRBs (purple lines), while the lower section exhibits the $\Gamma$ evolutions in SGRBs (red lines) and LGRBs (yellow lines). In both sections, solid lines depict the results considering the energy injections, whereas dashed lines represent the results without the energy injections. Furthermore, the black dot-dashed line in the upper section denotes the photospheric radius of the disks, i.e., $R_{\rm ph,disk}$.

In Figure \ref{Figure3}, the evolution of $R$ reveals that EFSs with the energy injections penetrate the disk photospheres earlier than those without the injections. This effect is notably more pronounced in LGRBs compared to SGRBs, indicating that higher injection powers and longer injection durations expedite EFSs' breakout from the disk photospheres. In addition, the $\Gamma$ evolution reveals that in some cases with relatively low disk densities and relatively small radii of the disk photospheres, the EFSs remain at relativistic velocities after breaking out of the disk photospheres due to the energy injections, such as the EFS with the energy injections in LGRBs of case 2 maintaining the Lorentz factor $\Gamma \sim 3$. However, these cases constitute a minority among all cases. In the majority, despite the energy injections, EFSs still exit the photospheres at non-relativistic velocities due to the hindrance posed by the high-density disk material. This suggests that, in most cases, higher injection powers and longer durations than those considered in this study would be necessary for EFSs to still maintain relativistic velocities after leaving the photospheres.

There are two exceptional cases, cases 1 and 9. In case 1, EFSs in LGRBs have initial radii outside the disk. In our study, outside the disk, the assumed environment is the ISM. Consequently, LGRBs in case 1 belong to typical LGRBs. Hence, we can observe that the corresponding $\Gamma$ exhibits significantly larger values in the initial phase and decays more slowly over time, in contrast to SGRBs (within the same case) where EFSs form inside the disk and their velocities decrease rapidly as they propagate in the high-density disk. In case 9, we can firstly see in Table \ref{table1} that the photospheric radius is 0. This indicates that the corresponding disk region is optically thin and low-density. Thus, in this case, there is no the RHDM, and the afterglows would appear earlier and brighter compared to those with EFSs formed in the optically thick regions. In addition, due to a low disk density, $\Gamma$ reach about 20 for SGRBs and 30 for LGRBs in the initial phase, much larger than those in cases 7 and 8 with higher disk densities.

\begin{figure*}[!t]
\centering
\includegraphics[width=1.0\linewidth]{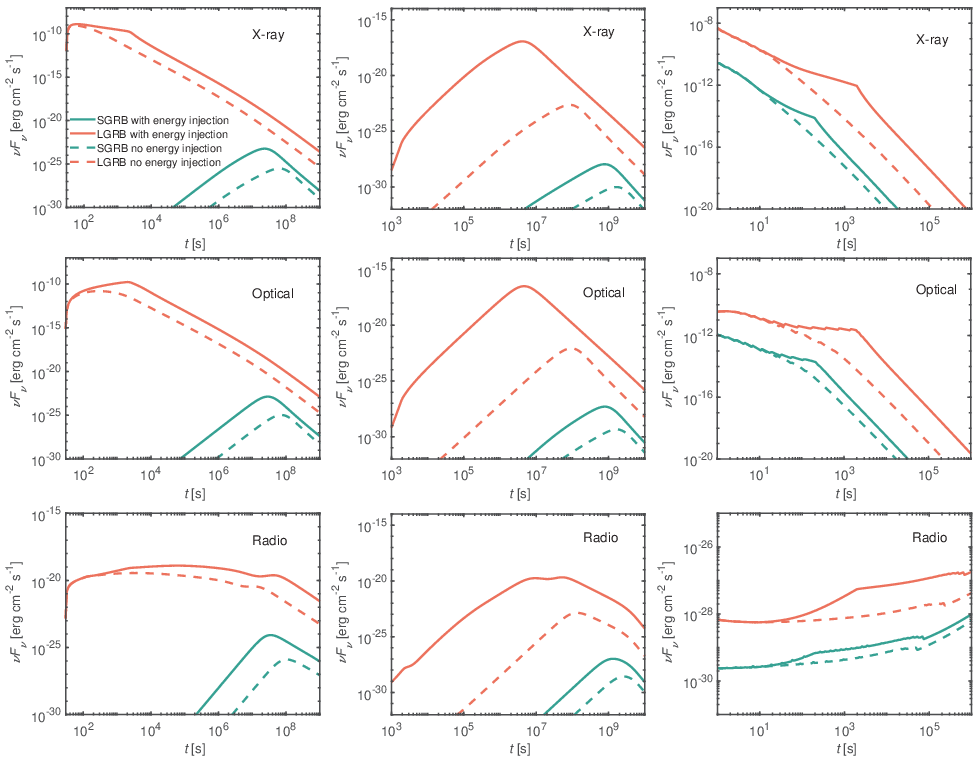}
\caption{Afterglow light curves for cases 1 (first column), 2 (second column), and 9 (third column). The first, second, and third rows show the X-ray, optical, and radio light curves, respectively. Green and red lines represent the results of SGRBs and LGRBs, where solid and dashed lines denote the results with and without the energy injections, respectively.}
\label{Figure5}
\end{figure*}

\subsection{Light curves}

There are two types of radiation discussed in our work. As mentioned in Section 2, one is the RHDM and the other is the afterglow synchrotron radiation from EFSs after they break out of the disk photospheres. Here we show their light curves.

\subsubsection{RHDM}

Figure \ref{Figure4} shows the RHDM optical ($5\times10^4 \,{\rm Hz}$) light curves. The purple and red lines denote the light curves in SGRBs, and the green and yellow lines represent the light curves in LGRBs. Meanwhile, the solid and dashed lines exhibit the results with and without the energy injections, respectively.

In the figure, the light curves with and without the energy injections are overlapped in all cases, indicating that the effects of the energy injections caused by the reactivation of GRB central engines are minimal for this type of radiation. The reason is that the observed luminosity from the shell layers heated before EFSs received the energy injections dominates the total observed luminosity, compared to the observed luminosity from the shell layers heated during EFSs received the energy injections, this is because the Lorentz factors of EFSs have been decayed to relatively lower values in the energy injection phase than in the initial formation phase due to the resistance of the high-density disk material. In addition, the fluxes of the light curves do not obviously change with the observer time. This is attributed to the fact that the densities of the disk material are large enough to lead to significantly long diffusion times and thus to observed luminosities much smaller than the injection luminosities, i.e., $L_{\rm ph} \ll L_{\rm inj}$, which means that the energy of the heated disk material cannot be effectively radiated. Moreover, since the EFSs form outside the disk, there is no the RHDM in the LGRB scenario of case 1.

It can be found that the fluxes improve significantly overall with the increasing radial radius by comparing the cases with the same SMBH masses but different radial radii, such as case 1 ($10^6 M_\odot$, $10^3 R_{\rm g}$), case 2 ($10^6 M_\odot$, $10^4 R_{\rm g}$), and case 3 ($10^6 M_\odot$, $10^5 R_{\rm g}$), which indicating RHDM fluxes are directly proportional to the radial radius of the disks. The reason is that, given the model parameter values, the larger the radial radius, the larger the photospheric radius of the disks (before about $R_{\rm r}<10^5 R_{\rm g}$), which hints that EFSs break out of the disk photospheres with more time, and thus more energy from the EFSs is absorbed by the disk material.

\subsubsection{Afterglows}

Figure \ref{Figure5} shows the afterglow light curves at X-ray ($1 \, {\rm keV}$), optical ($5 \times 10^{14} \, {\rm Hz}$), and radio ($10^9 \, {\rm Hz}$) bands for cases 1, 2, and 9. The top, middle, and bottom rows represent the X-ray, optical, and radio light curves, respectively. The green and red lines depict the light curves for SGRBs and LGRBs, while the solid and dashed lines differentiate the results with and without the energy injections.

In Figure \ref{Figure5}, light curves with the energy injections exhibit notable distinctions in both profiles and peak fluxes compared to those without, suggesting evident effects of the energy injections on the afterglow light curves. This stands in contrast to the RHDM.

In case 1, the light curves in LGRBs peak considerably earlier and exhibit significantly larger peak fluxes compared to those in SGRBs. This is because, in this case, EFSs in LGRBs form outside the disk, whereas EFSs in SGRBs form inside the disk and have become unrelativistic as they break out of the disk photosphere. In case 2, although both the EFSs in SGRBs and LGRBs form inside the disk, the EFS in LGRBs, supported by the energy injections, maintains relativistic velocities after breaking out of the disk photosphere (as seen in Figure \ref{Figure3}). Consequently, an earlier peak time and a larger peak flux are observed in this scenario compared to that without the energy injections in LGRBs and to those in SGRBs. In case 9, the disk reaches an optically thin state, allowing the afterglow radiation to nearly insusceptibly pass out of the disk. However, significantly strong synchrotron self-absorption occurs in the radio wavelengths when the radiation is produced by the EFSs so that the radio fluxes are markedly lower. In addition, also due to the higher synchrotron self-absorption, the flux gap between the X-ray and optical bands is larger than that in case 1 with the afterglows produced outside the disk and than that in case 2 with the afterglows contributed by the EFSs after the breakout from the disk photospheres. Finally, the light curves in the other cases, regardless of whether they are in LGRBs or SGRBs, resemble those of SGRBs in cases 1 and 2, i.e., rather weak afterglows. Therefore, we have not included the afterglow light curves for these cases here.

\section{Conclusions and discussion}

In this paper, we focus on the effects of energy injections supported by the reactivation of GRB central engines on the dynamical evolution of EFSs propagating in AGN disks with a Gaussian density profile in the vertical direction, as well as on the light curves of two types of radiation, the RHDM and the afterglows. As results, we find that EFSs with the energy injections can break out of the disk photospheres faster than those without, and the EFSs with the energy injections in LGRBs occurring at locations with low densities and small photospheric radii still retains relativistic velocities after penetrating the disk photospheres. Furthermore, we find that the light curves with and without the energy injections are not distinct for the RHDM but are notable differences in peak fluxes for the afterglows.

\begin{figure}
\centering
\includegraphics[width=0.95\linewidth]{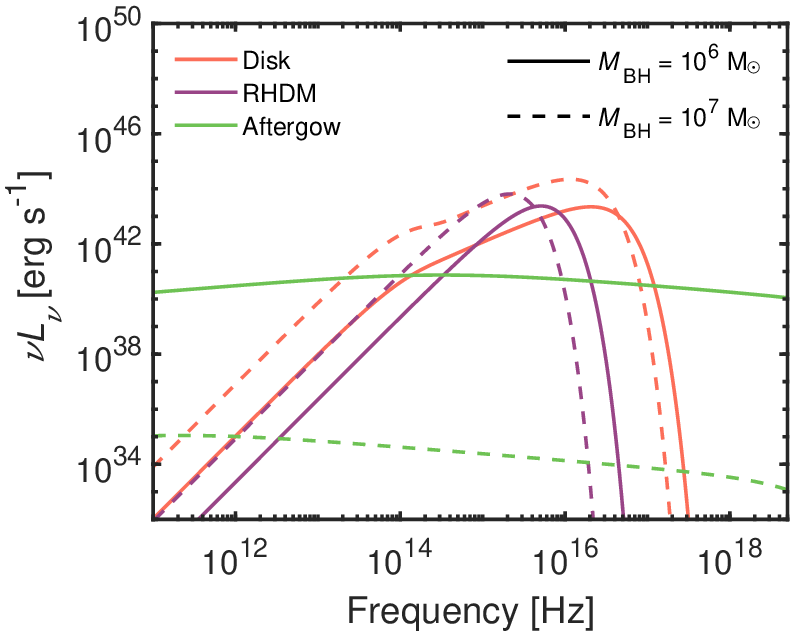}
\caption{Spectral energy distributions of the disk, RHDM, and synchrotron afterglows for $M_{\rm BH} = 10^6, 10^7 \, M_\odot$, $l_{\rm E}=0.5$, and $\alpha=0.01$, where solid and dashed lines denote the results with $M_{\rm BH} = 10^6 \, M_\odot$ and $10^7 \, M_\odot$, respectively.}
\label{Figure6}
\end{figure}

A hollow cone would appear behind the EFS as an EFS driven by an initially ejected GRB jet sweeps up the disk material, and the hollow cone should gradually fill with the heated disk material. If the hollow cone is completely filled before new ejected jets catch up with the EFS, there would be no energy injections. Thus, here we estimate the time scale of the fully filled hollow cone to evaluate the rationality of the adopted start time of energy injections. This time scale can be estimated by $t_{\rm c}\approx R_{\rm c}/\upsilon_{\rm c}$, where $R_{\rm c}$ is the inner radius of the cross section of the heated disk material, and $\upsilon_{\rm c}=P_{\rm c}/\rho_{\rm disk}$ is the lateral expansion velocity \citep{Begelman1989,Bromberg2011}. $P_{\rm c}=E_{\rm c}/2V_{\rm c}$, $E_{\rm c}=\int_{t_{\rm s}}^{t_{\rm e}} L_{\rm inj} dt_{\rm b}$, and $V_{\rm c}=\pi R_{\rm e}^2 (R_{\rm e}-R_{\rm ES})(\tan^2\theta_{\rm c}-\tan^2\theta_{\rm jet})$ are the pressure, energy, and volume of the heated disk material, respectively. $t_{\rm s}$ is the time of EFS formation in the local rest frame, ${t_{\rm e}}$ is the start time of the energy injections in the local rest frame, and $R_{\rm e}$ is the location of the EFS at the start time. In this work, the start time of energy injections is set to $10\,{\rm s}$ in the observer frame, i.e., $T_{\rm st,in}=10 \,{\rm s}$. Using the SGRB with the energy injections in case 1 for an example, one have ${t_{\rm s}}\approx 25 \, {\rm s}$, ${t_{\rm e}}\approx 67 \, {\rm s}$, and $R_{\rm e}=1.8\times10^{12} \,{\rm cm}$. Further, we obtain the time scale of the fully filled hollow cone $t_{\rm c}\approx 98 \,{\rm s}$, which is larger than the start time of the energy injections (${t_{\rm e}}\approx 67 \, {\rm s}$), indicating that setting $T_{\rm st,in}=10 \,{\rm s}$ is reasonable.

Using the same disk parameters as mentioned in Section 2.1, we estimate the spectral energy distributions of the disk model \citep{Goodman2003}, RHDM, and synchrotron afterglows, and compare the optical ($5\times 10^{14}\,{\rm Hz}$) luminosity from the RHDM and afterglows and the X-ray ($1\,{\rm keV}$) luminosity from the afterglows with the corresponding disk luminosity. Regarding the RHDM, the optical luminosity at smaller radial radii is overshadowed by the optical luminosity of the disk model. However, at larger radial radii, the optical luminosity can outshine that of the disk model. For example, in case 3, the optical luminosity of the RHDM reaches $2.4\times10^{43}\,{\rm erg\,s^{-1}}$, exceeding the corresponding disk luminosity values of $8.3\times10^{42}\,{\rm erg\,s^{-1}}$. Although the optical luminosity of the RHDM outshines that of the disk model at larger radial radii, they may be indistinguishable due to two factors: (1) the luminosity excess is not markedly obvious, and (2) the diffuse timescale is sufficiently large, causing the optical luminosity of the RHDM not to decay significantly over relatively short times. Regarding the afterglows, the optical luminosity is encompassed by the optical luminosity of the disk model in optically thick regions. However, it can surpass the corresponding disk luminosity in optically thin regions (e.g., case 9) or when EFSs form outside the disks (e.g., the LGRBs in case 1). The X-ray luminosity scenario is similar to the optical luminosity scenario. But, if EFSs still maintain relativistic velocities after breaking out of the disk photospheres, such as the LGRB with the energy injections in case 2, the X-ray luminosity can exceed the corresponding disk luminosity. Here, in contrast to the RHDM, the afterglows can be easily distinguished from the disk radiation due to their relatively short durations. Finally, as an example, Figure \ref{Figure6} shows the spectral energy distributions of the disk model (red), RHDM at $R_{\rm r}=10^5 \, R_{\rm g}$ (purple), and synchrotron afterglows at $R_{\rm r}=10^4 \, R_{\rm g}$ (green) for $M_{\rm BH} = 10^6, 10^7 \, M_\odot$, $l_{\rm E}=0.5$, and $\alpha=0.01$. It can be seen that the spectral energy distributions of the afterglows are significantly broader than those of the disk model and RHDM. However, for the same accretion rate, the afterglow luminosities decrease dramatically with the SMBH masses increasing. They are well below the luminosity of the AGN disk, which implies that it is more difficult to observe the afterglows for the more massive SMBHs.

Moreover, different values of parameters of accretion system can significantly affect the above results. For the given accretion rate and viscosity parameter, the SMBH mass has inverse correlation with the disk densities and positive correlation with the disk half-thicknesses and luminosity \citep[e.g.,][]{Kato2008book}. This implies that EFSs require more time to penetrate the disk photosphere for larger SMBH mass. Similarly, the accretion rate is inversely proportional to disk density and proportional to disk half-thickness and luminosity. In contrast, the viscosity parameter is inversely proportional to disk density, half-thickness, and luminosity. Therefore, a low-mass SMBH, the low accretion rate, as well as the large viscosity parameter, could favor the appearance of GRB afterglows.

In this work, the assumed durations of energy injections in SGRBs and LGRBs are referenced to times obtained from GRBs occurring in typical environments. However, in the AGN disk, the long-duration fallback accretion to the central BH might be maintained to prolong the timescale of the energy injection. Besides, if cavities with radii close to the disk half-thickness are involved \citep{Kimura2021,Yuan2022}, EFSs will encounter less deceleration while propagating within these cavities, enabling them to maintain relativistic velocities as they escape from the disk photospheres, which suggests the anticipation of stronger afterglows.

\acknowledgments
We thank Bing Zhang, Li Xue, Da-Bin Lin, and Yan-Qing Qi for helpful discussions. This work was supported by the National Key R\&D Program of China (Grant No. 2023YFA1607902), the National Natural Science Foundation of China (Grant Nos. 12173031, 12221003, and 12303049), and the National Fund for Postdoctoral Researcher Program (Grant No. GZC20231424).

\end{document}